\documentclass[conference]{IEEEtran}
\IEEEoverridecommandlockouts
\usepackage{cite}
\usepackage{amsmath,amssymb,amsfonts}
\usepackage{algorithmic}
\usepackage{graphicx}
\usepackage{textcomp}
\usepackage{amsmath}
\usepackage{xcolor}
\usepackage{soul}
\usepackage{enumitem}
\usepackage[pscoord]{eso-pic} 
\def\BibTeX{{\rm B\kern-.05em{\sc i\kern-.025em b}\kern-.08em
    T\kern-.1667em\lower.7ex\hbox{E}\kern-.125emX}}
\begin{document}

\newcommand{\placetextbox}[3]{ 
    \setbox0=\hbox{#3} 
     \AddToShipoutPictureFG*{ 
     \put(\LenToUnit{#1\paperwidth},\LenToUnit{#2\paperheight}){ 
      \vtop{{\null}\makebox[0pt][c]{#3}}} 
    } 
} 
\placetextbox{.2}{0.055}{ 978-3-903176-44-7~\copyright~2022 IFIP}

\title{Planning a Cost-Effective Delay-Constrained Passive Optical Network for 5G Fronthaul
}
\author{\IEEEauthorblockN{Abdulhalim Fayad, Manish Jha, Tibor Cinkler,  and Jacek Rak }
\thanks{Abdulhalim Fayad, Manish Jha, and Tibor Cinkler are with Department of Telecommunications and Media Informatics, Budapest University of Technology and Economics, Hungary.
Jacek Rak and Tibor Cinkler are with Department of Computer Communications, Gda\'nsk University of Technology, Gda\'nsk, Poland.
 e-mail: \{Fayad, Manish, Cinkler\}@tmit.bme.hu, jrak@pg.edu.pl}}
\maketitle
\begin{abstract}
With the rapid growth in the telecommunications industry moving towards 5G and beyond (5GB) and the emergence of data-hungry and time-sensitive applications, Mobile Network Operators (MNOs) are faced with a considerable challenge to keep up with these new demands. Cloud radio access network (CRAN) has emerged as a cost-effective architecture that improves 5GB performance.
The fronthaul segment of the CRAN necessitates a high-capacity and low-latency connection. Optical technologies presented by  Passive Optical Networks (PON) have gained attention as a promising technology to meet the fronthaul challenges. In this paper, we proposed an Integer Linear Program (ILP) that optimizes the total cost of ownership (TCO) for 5G  using CRAN architecture under different delay thresholds. We considered the Time and Wavelength Division Multiplexing Passive Optical Network (TWDM-PON) as a fronthaul with different splitting ratios. 
\end{abstract}

\section{Introduction}
 The emergence of high bandwidth-demanding and stringent latency requirements applications such as augmented reality, sophisticated online video gaming, security applications, intelligent farming, connected vehicles, as well as the exponential growth of mobile traffic, which is expected to exceed 5000~EB/month by 2030\cite{b1}, pose significant challenges to mobile network operators (MNOs). Therefore academia and industry started researching 5GB mobile technologies to overcome the increase in the number of end-user demands\cite{b2,b3}. One of the  innovative solutions to improve the performance of the networks cost-effectively is cloud radio access networks (CRAN)\cite{b4}. The processing operations are carried out at the baseband unit (BBU) positioned in a central location in the CRAN architecture. The remote radio heads (RRHs), on the other hand, are located on the antenna side and have a limited range of responsibilities\cite{b4}. The fronthaul network is a part of CRAN that links  RRHs with their serving BBU, and it has a high need for high-bandwidth, low-latency connection\cite{b5}. However, the cost of the fronthaul is a challenge, although the C-RAN can lower both operational and capital costs (Opex and Capex).
 
\begin{figure}[!htb]
\centering
\includegraphics[width=0.42\textwidth]{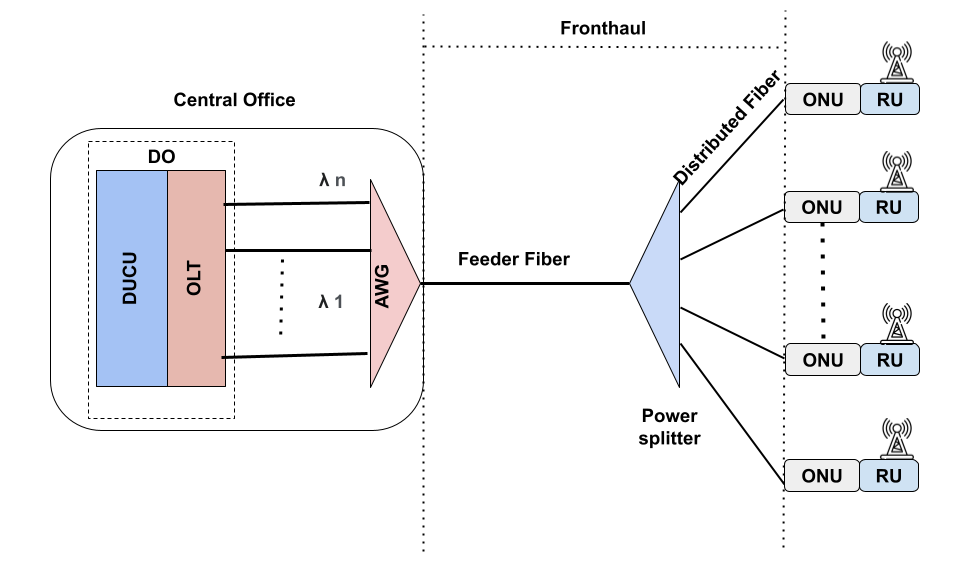}
\caption{C-RAN architecture with TWDM-PON fronthaul}
\label{fig:2}
\end{figure}

 Many technologies have been proposed for the fronthaul architecture, such as Free Space Optics, millimeter-waves,  microwave, fiber optics, and optical access networks \cite{b9}. Optical access networks presented by Passive Optical Networks (PON) play an essential role behind the success of 5GB. They have a point-to-multipoint topology, which allows for effective use of fiber resources. Moreover, they provide low latency and support a massive volume of data traffic\mbox{\cite{b7,b8}}. TWDM-PON (time- and wavelength-division multiplexed passive optical network) has received much attention for next-generation optical access systems. One of its primary uses is to handle mobile fronthaul streams that demand low latency and high capacity\cite{b10}.  
 Recently, the 3GPP, and the IEEE~WG~1914 recommended redesigning the two layers  C-RAN (BBU and RRH) architecture into a new design that defines three baseband function layers: central unit (CU), distributed unit (DU), and a remote unit (RU)\cite{b8,b11}. There are four potential scenarios  to deploy 5G networks \cite{b8}. 
 Deployment scenario 1 is most suitable for latency-sensitive services in terms of both low-latency communications and actual implementation costs.\\ In this paper, we focus on designing a cost-effective optical fronthaul based on TWDM-PON architecture for  delay-sensitive services  that  have a strict requirement for low latency connection, which is a critical problem in terms of enhancing the  quality of service (QoS)\cite{b6}.
 Our proposed C-RAN based on TWDM-PON fronthaul architecture following  deployment scenario 1  is illustrated in Fig.~\ref{fig:2}.

This paper is organized as follows. Section II reviews the related works. Section III describes the studied problem. The problem is formulated in Sections IV-V. A case study, and numerical findings are presented in Section VI. Finally, Section VII concludes the paper.
 \section{Related Works }
There has been a lot of focus in recent years on designing \mbox{a cost-effective} and low latency fronthaul for 5G. In \cite{b12}, Wang, Xin et al. introduce a MILP model and a heuristic approach to reduce the joint cost of latency and network deployment of a TDM-PON based MEC-enabled C-RAN. Ranaweera et al. in \cite{b13} analyze alternative optical fronthaul networks for 5G \mbox{C-RAN} design to produce a low-latency, bandwidth efficient, and cost-effective fronthaul network in classical CRAN. Masoudi, Meysam et al.\cite{b14} propose an Integer Linear Program (ILP) and a genetic algorithm to minimize the total cost of ownership (TCO) of CRAN, and assess the cost of migrating to a C-RAN-based \mbox{TWDM-PON} architecture with full network function centralization and partial centralization using function splitting. Ranaweera et al. in \cite{b15} propose a~generalized joint-optimization framework based on ILP for 5G FWA networks that simultaneously plans wireless access and optical transport while fulfilling diverse network constraints. In \cite{b16}, Wang, Nan et al. provide a~low-delay layout planning CRAN employing WDM-PON as \mbox{a~fronthaul}. In addition, the planning process is presented using the nonlinear decreasing inertia weight particle swarm optimization technique \mbox{(NL-wPSO)}. 
Marotta et al.  \cite{b17} propose an ILP model to evaluate the optimal deployment of 5G C-RAN fronthaul  using point to optical fiber and microwave links under delay constraints in a~brownfield scenario.
However, most of the existing studies do not consider the planning of the CRAN fronthaul deployment under different delay thresholds. Also, they do not analyze how the delay values can influence the total cost of ownership (TCO)  of the network that can help the MNOs plan their networks to be ready for upcoming time-sensitive services.

\section{Problem Description}
Designing a cost-effective 5G CRAN fronthaul based on the proposed architecture in Fig.~\ref{fig:2} for time-sensitive services can be stated as follows. Given all RU/ONU locations, all potential locations for splitters and the potential locations for the central offices (COs),
each RU/ONU (RU and ONU are co-located) can connect to its power splitter by the distribution fiber, and each splitter can connect to the central office by the feeder fiber. Each central office contains a number of  DUCUs (DU co-located with CU), many optical line terminals (OLTs), where DUCU and OLT are co-located, as well as a number of arrayed waveguides (AWGs) that connect the DO (we will use the abbreviation DO to express the DUCU and OLT together). 
Our optimization problem is to find the optimal locations of the COs and power splitter to find the shortest path from each RU/ONU towards the central office, meeting the delay constraints for different splitting ratios. As a result, this leads to minimizing the TCO of the networks based on the delay threshold. The total delay over the CRAN fronthaul network $T_{f}$ can be calculated by reformulating the equation as given in \cite{b12}:
\begin{equation}
T_{f}=T_{s}+T_{q}+T_{co}+T_{d}
\end{equation}
where $T_{s}$ refers to the required time to $send$ the data from each  RU/ONU  to CO; $T_{q}$ is referred to the $queuing$ delay;  $T_{co}$ is the delay caused by data $processing$ in each CO; $T_{d}$ is the $propagation$ delay in optical fiber. $T_{s}$, 
$T_{q}$, and $T_{co}$ are related to the hardware of the network, so that all of them will be ignored. We only consider the propagation delay $T_{d}$ as it is considered the main bottleneck of the one-way transmission latency, and   it can be calculated as follows:
\begin{equation}
    T_{d}= T_{d1}+ T_{d2}= \alpha \cdot(d1+d2)
\end{equation}
where $T_{d1}$, and $T_{d2}$ are the propagation delay over feeder fiber and distribution fiber respectively. $d1$, and $d2$ are the length of feeder fiber and distribution fiber. $\alpha$  is the propagation delay per kilometer of fiber which is 5 $\mu s$; The maximum propagation delay over CRAN is 50 $\mu s$ \cite{b18}.

\section{Total Cost of Ownership Modeling}
This section provides a cost model for TCO that covers Capex and Opex of 5G CRAN fronthaul based on TWDM-PON architecture. In our study, we consider that only one operator can serve that studied area. Furthermore, there is no infrastructure sharing, and all equipment and infrastructure are related to that operator and no need for leasing fiber. Therefore, TCO can be calculated as follows by reformulating the model presented in \cite{b19}:
\begin{equation}
  TCO=Capex+N_{r}\cdot Opex  
\end{equation}
where $N_{r}$ is the number of years.
\vspace{2 mm}
\subsubsection{\textbf{Capex}}
The term ``Capex'' refers to a one-time investment expense for acquiring or upgrading physical assets or infrastructure. Our approach takes into account the cost of equipment, infrastructure, and installation:
\begin{equation}
    Capex=Eq_{cost}+Inf_{cost}+Ins_{cost}
\end{equation}
\begin{enumerate}[label=\Alph*]
    \item Equipment costs:
This refers to all costs associated with purchasing equipment for the 5G CRAN architecture:
\begin{equation}
\begin{aligned}
  Eq_{cost}=N_{DO}C_{DO}+N_{AWG}C_{AWG}\\+N_{PS}C_{PS}+N_{RU/ONU}C_{RU/ONU}
\end{aligned}
\end{equation}
where $N_{DO}$, $C_{DO}$, $N_{AWG}$, $C_{AWG}$, $N_{PS}$, $C_{PS}$, $N_{RU/ONU}$ and $C_{RU/ONU}$ denote the number and the cost for each of DOs, AWGs, power splitters and RU/ONUs  repectively.
\vspace{2 mm}
\item Infrastructure costs:
This refers to the overall cost of deployment. Because the length of the fiber determines the length of the duct and trenching, we have linked the fiber cost to the infrastructure component.
\begin{equation}
    Inf_{cost}=d(C_{f}+C_{cw})
\end{equation}
where $d$ denotes the length of fiber cable. $C_{f}$ and $C_{cw}$ refer to fiber optic cable cost and civil work respectively.
\vspace{1 mm}
\item Installation costs: installation man-hours, wiring, site preparation, technician remuneration, and travel time to and from site locations are all included in the installation component.
\begin{equation}
    Ins_{cost}=\left[\sum_{i=1}^{N_{\text {link }}}\left( T_{i}+2 T_{t}\right) \cdot  TS\right] TN
\end{equation}
where $T_{i}$ and $T_{t}$ denote installation time and travel time, $TS$ and $TN$ represent
the technician salary and the number of required technicians respectively.
\vspace{2 mm}
\end{enumerate}
\subsubsection{\textbf{Opex}} Opex means operational expenditures,  which refers to the ongoing costs of operating the network on a daily basis. Energy consumption ($C_{E}$), operation and maintenance ($C_{OM}$) and site rental cost ($C_{Sr}$) are the three key Opex components. Opex can be calculated as follows:
\begin{equation}
    Opex=C_{E}+C_{OM}+C_{Sr}
\end{equation}
\begin{enumerate}[label=\Alph*]
    \item Energy consumption: Access networks are projected to consume 70$\%$ of the total energy consumed by telecommunication networks\cite{b20}. It can be calculated by adding the consumption costs of all electrical equipment in various locations of the network as follows:
\begin{equation}
\footnotesize
C_{E}=E_{p}\cdot365\cdot24\cdot\left(\sum_{i \in N} (C^{EDO}_{i}+C^{Ecool}_i)+\sum_{i \in M}C^{ER}_{i}\right)
\end{equation}
where $E_{p}$, $C^{EDO}$, $C^{Ecool}$ and $C^{ER}$ represent the cost of power bills for one Wh energy consumption  per hour in DO, cooling system, and RU/ONU, respectively, while $N$ and $ M$ are the number of DOs, and RU/ONUs, respectively.
\vspace{2 mm}
\item Operation and maintenance: denotes the regular maintenance program required to keep the network up and running. This includes equipment monitoring and testing, software upgrades (including license renewals as needed),  battery replacement and the annual operation and maintenance costs are equal to 10$\%$ of Capex\cite{b19}.
\begin{equation}
\begin{split}
& C_{O\&M}=  \sum_{i \in N} (C^{MDO}_i)+\sum_{i \in nf}C^{MR}_i+ S_{lic}
\end{split}
\end{equation}
where $C^{MDO}$ and $C^{MR}$ represent the operation and maintenance costs for  DO, OLTs and RU/ONU, respectively, while $S_{lic}$ denotes software upgrade cost.
\vspace{2 mm}
\item Site rental: this aspect refers to the price that mobile network operators pay to rent space for their equipment on an annual basis\cite{b20}, which can be calculated as:
\begin{equation}
    C_{Sr}=N \cdot Sr_{y}
\end{equation}
where $N$, and $Sr_{y}$ denote the number of cell sites and yearly costs for one cell site rental, respectively.
\end{enumerate}
\section{ILP formulation}
Given a set of RU/ONUs locations, and a set the potential locations of COs, the goal is to find the optimal number and location of Central offices, DOs, AWGs and splitters to obtain the optimal deployment costs of the network under the delay constraints. Each RU/ONU is connected to a splitter by a~distributed fiber. The splitter is connected to the feeder fiber to the AWG, and then each AWG is connected to the DO. We consider that the DO and the AWG are located at the central office. This paper proposes an Integer Linear Program (ILP)  that can be applied in any given scenario to find the optimal locations of splitters and central offices and the optimal paths between RU/ONUs, splitters, and central offices.
\subsubsection{Network data sets and parameters}
In our framework, we employ several datasets to represent various network locations. Table~\ref{tab:1} lists these sets and their descriptions. The framework is also built around a set of parameters that can be tweaked to fit the deployment scenario. Table~\ref{tab:2} summarizes the parameters. 
\begin{table}
\caption{Network Data Sets}
\label{tab:1}       
\begin{tabular}{p{1.5cm}|p{6.5cm}}
\hline\noalign{\smallskip}
Notation & Description  \\
\noalign{\smallskip}\hline\noalign{\smallskip}
$C$  & Set of potential locations for the Central Offices (COs) where DOs, and  AWGs are located \\
$S$ & Set of potential locations for splitters  \\
$R$ & Set of potential locations for RUs/ONUs   \\

\noalign{\smallskip}\hline
\end{tabular}
\end{table}

\begin{table}
\caption{Network Parameters}
\label{tab:2}       
\begin{tabular}{p{1.5cm}|p{6.5cm}}
\hline\noalign{\smallskip}
Notation & Description  \\
\noalign{\smallskip}\hline\noalign{\smallskip}
$N_{C}$ & Number of central offices\\
$N_{DO}$ & Number of DOs (DUCU + OLT)\\
$N_{a}$ & Number of AWGs\\
$N_{s}$ & Number of splitters\\
$N_{r}$ & Number of RU/ONUs\\
$d_{ij}$ & The distance between the $i^{t h}$ CO and the $j^{t h}$ splitter (feeder fiber)\\
$d_{jr}$ & The distance between the $j^{t h}$ splitter and the $r^{t h}$ RU/ONU (distributed fiber)\\
$d1_{\max}$  & The maximum allowed distance between each RU/ONU and the power splitter (distributed fiber) \\
$d_{\max}$  & The maximum allowed distance between each RU/ONU and the central office \\
$H$& Maximum number of DOs in the central office\\
$\eta$ & Splitting ratio {1:4, 1:8, 1:16}\\
$C_{F}$ & The cost of fiber optic cable per meter\\
$C_{s}$ & The cost of a splitter\\
$C_{r}$ & The cost of RU/ONU\\
$C_{DO}$&  The cost of DO\\
$C_{Co}$&  The cost of central office\\
$C_{a}$& The cost of AWG\\
 $\tau$ &  Fiber optic cable propagation delay\\
 $\tau_{\max }$ &  Maximum allowed  fronthaul propagation delay between the RU/ONU  and the central office \\
 $\theta_{D}$ & 
Passive optical network downlink capacity\\  
$\theta_{U}$ & 
Passive optical network uplink capacity\\ 
 $\theta_{rd}$ & RU/ONU downlink capacity \\
 $\theta_{ru}$ & RU/ONU uplink capacity \\
\noalign{\smallskip}\hline
\end{tabular}
\end{table}
\subsection{Decision Variables}
\begin{itemize}
\item Binary variable $x_{ij}$\\
$x_{ij}$=$\left\{\begin{array}{l}1 \textrm { if the  $i^{th}$ CO and the $j^{th}$ splitter are connected}  \\ 0 \textrm { otherwise }\end{array}\right.$
\item Binary variable $x_{jr}$\\
$x_{jr}$=$\left\{\begin{array}{l}1 \textrm {  if  the $j^{th}$ splitter  and $r^{th}$ RU/ONU are connected}  \\ 0 \textrm { otherwise }\end{array}\right.$
\item Binary variable $C_{i}$\\
$C_{i}$=$\left\{\begin{array}{l}1 \textrm { if the $i^{th}$ CO is selected } \\ 0 \textrm { otherwise }\end{array}\right.$
\item Binary variable $S_{j}$\\
$S_{j}$=$\left\{\begin{array}{l}1 \textrm { if  the $j^{th}$ splitter is selected} \\ 0 \textrm { otherwise }\end{array}\right.$
\item Binary variable $R_{r}$\\
$R_{r}$=$\left\{\begin{array}{l}1 \textrm { if  the $r^{th}$ RU/ONU is selected} \\ 0 \textrm { otherwise }\end{array}\right.$
\end{itemize}
\vspace{5 mm}
\subsection{Objective Function}
\begin{equation}
\vspace{-1.2 cm}
\begin{split}
\hspace{-0.5 cm}min\quad \underbrace{C_{Co} N_{C} }_{\textrm {CO cost}}+\underbrace{C_{DO} N_{DO} }_{\textrm {DO cost}}+\underbrace{{C}_{\mathrm{a}} N_{a}}_{\textrm{AWGs cost}}+ \underbrace{{C}_{\mathrm{s}}\sum_{j=1}^{N_{s}} S_{j}}_{\textrm{splitters cost}}+\underbrace{{C}_{\mathrm{a}}\sum_{r=1}^{N_{r}} R_{r}}_{\textrm{RU/ONUs cost}}\\+\underbrace{ C_{F}\sum_{i=1}^{N_{\mathrm{DO}}} \sum_{j=1}^{N_{\mathrm{s}}} \sum_{r=1}^{N_{\mathrm{r}}} (x_{ij}d_{ij}+x_{jr}d_{jr})}_{\textrm {Fronthaul deployment cost}}+ \underbrace{{C}_{\mathrm{E}}+ C_{O\&M}+C_{Sr}}_{\textrm{Opex}}
\end{split}
\end{equation}  
\vspace{0.5 cm}
\subsection{Constraints}
\begin{enumerate}
\item Topology constraints
  \begin{enumerate}
  \item Each RU/ONU should be connected to only one splitter:\\
  \begin{equation}
  \sum_{j \in S} x_{jr}=1 \quad \forall r \in R
 \end{equation}
\item The split ratio of each splitter must not be exceeded by the number of RU/ONUs connected to it:
\begin{equation}
\sum_{r \in R} x_{jr} \leq \eta \quad \forall j \in S
\end{equation}
\item If there is an optical link from this splitter to an RU/ONU, it should be installed at a viable splitter location:
\begin{equation}
    x_{jr} \leq S_{j} \quad j \in S, r \in R
\end{equation}
\item If a splitter is used at a possible site, it must be connected to at least one RU/ONU:
\begin{equation}
\sum_{r \in R} x_{jr} \geq S_{j} \quad \forall j \in S    
\end{equation}

\item Each splitter should be connected to only one CO:
\begin{equation}
\sum_{i \in C} x_{ij}=1 \quad \forall j \in S
\end{equation}
\item The number of DOs in each central office can not exceed the maximum number:
\begin{equation}
N_{DO}= \sum_{i \in C} x_{ij}\leq H \quad \forall j \in S
\end{equation}
\item If an optical path exists between a DO and a splitter, the splitter must be connected to at least one RU/ONU:
\begin{equation}
  \sum_{i \in C} x_{ij}\leq \sum_{j \in S} x_{jr}  \quad \forall r\in R
\end{equation}
\item The number of splitters in the network should be equal to the total number of DOs:
\begin{equation}
  \sum_{j \in S} S_{j} =\sum_{j \in S} x_{ij}  \quad \forall i\in C 
\end{equation}
\item Number of AWGs in the network should be equal to the total number of splitters:
\begin{equation}
  N_{a} =\sum_{j \in S} x_{ij}  \quad \forall i\in C 
\end{equation}
\end{enumerate}
\item Capacity constraints
\begin{enumerate}
\item The capacity of downlink transmission must be equal or less than the maximum downlink capacity of TWDM-PON:
\begin{equation}
 \sum_{r \in R} \theta_{rd} x_{jr} \leq \theta_{D}  \quad \forall j\in S   
\end{equation}
\item The capacity of uplink transmission must be equal to or less than the maximum uplink capacity of TWDM-PON:
\begin{equation}
 \sum_{r \in R} \theta_{ru} x_{jr} \leq \theta_{U}  \quad \forall j\in S   
\end{equation}
\end{enumerate}

\item {Delay constraints:}
the fronthaul delay must not exceed the maximum allowed delay:
\begin{equation}
\tau (x_{ij}d_{ij}+x_{jr}d_{jr})\leq \tau_{max}\quad \forall i\in C, j \in S, r \in R
\end{equation}
\item Distance constraints
\begin{enumerate}
\item The maximum length of distribution fiber should not surpass the maximum specific length: 
\begin{equation}
 x_{ij}d_{ij} \leq   d1_{max}\quad \forall i\in C, j \in S
\end{equation}
 \item The distance between each RU/ONU and its serving central office must be less than the maximum distance allowed in PONs:
\begin{equation}
x_{ij}d_{ij}+x_{jr}d_{jr}\leq d_{max}\quad \forall i\in C, j \in S, r \in R
\end{equation}
\end{enumerate}

\end{enumerate}

\section{Case Study and Numerical Results}
\begin{figure}[!htb]
\centering
\includegraphics[width=0.5\textwidth,height=4.8cm
]{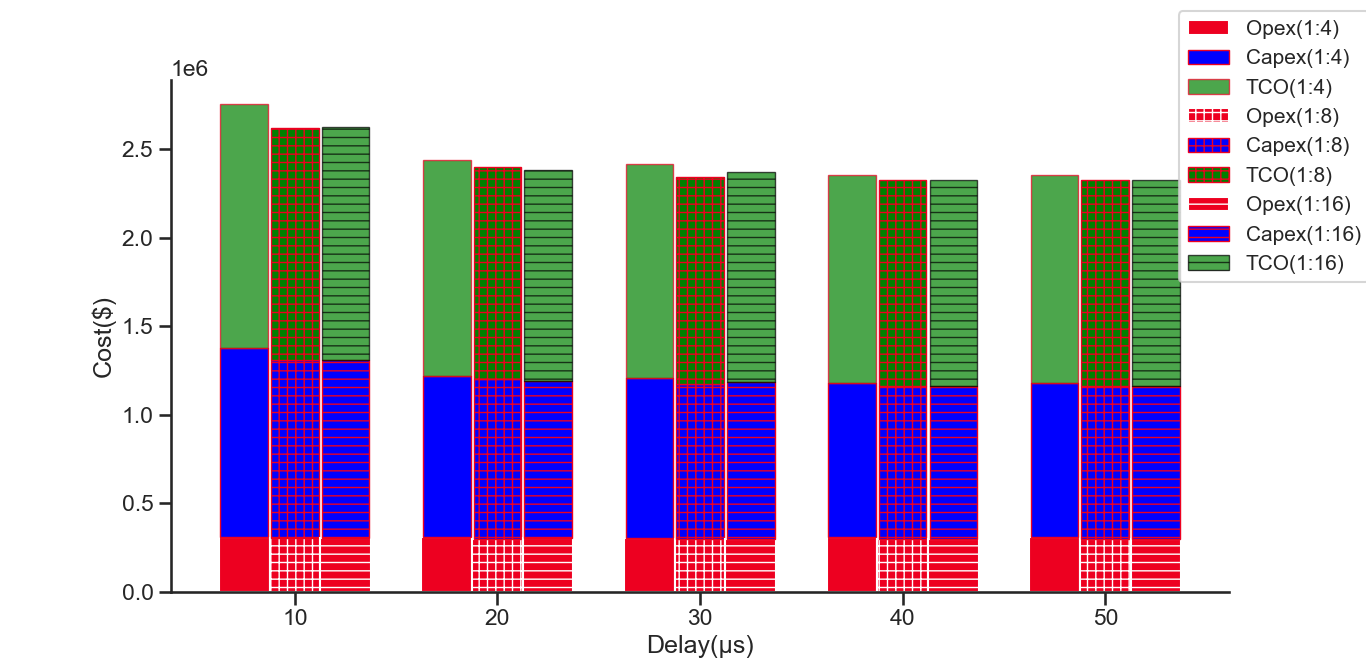}
\caption{TCO vs. delay threshold}
\label{fig:3}
\end{figure}

\begin{figure}[!htb]
\centering
\includegraphics[width=0.5\textwidth,height=4.8cm]{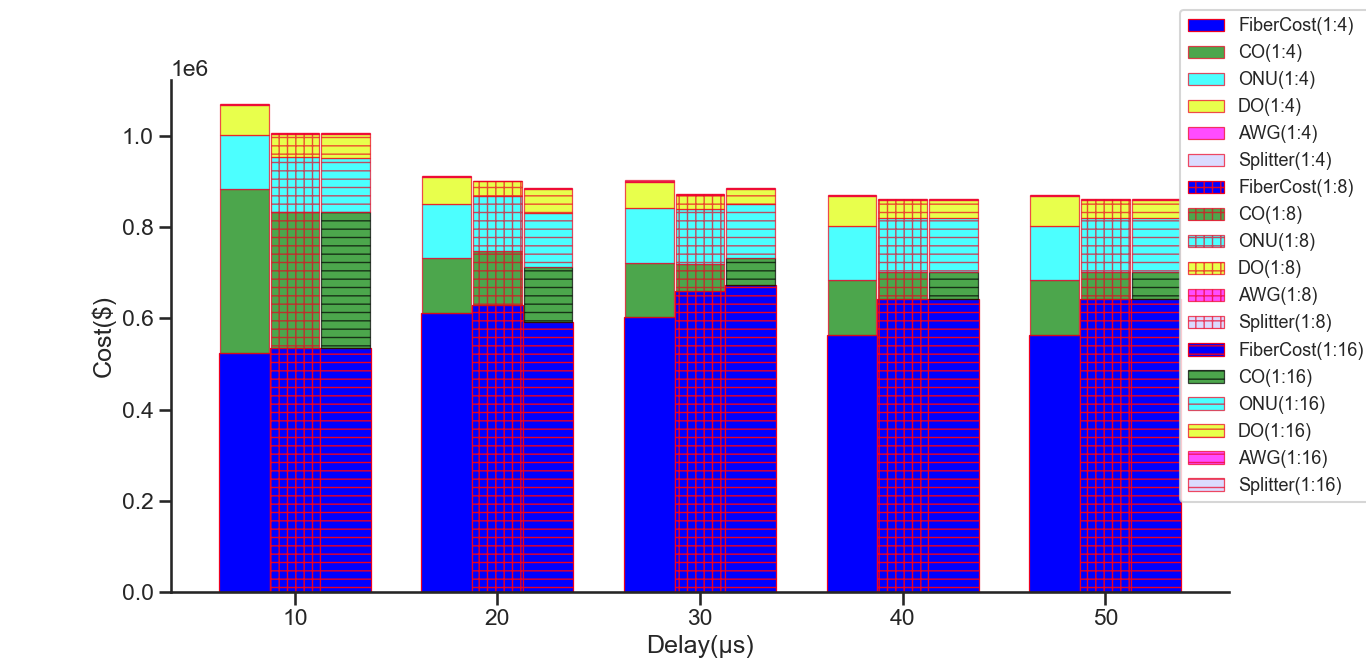}
\caption{Capex breakdown vs. delay threshold}
\label{fig:4}
\end{figure}

\begin{figure}[!htb]
\centering
\includegraphics[width=0.5\textwidth,height=4.8cm]{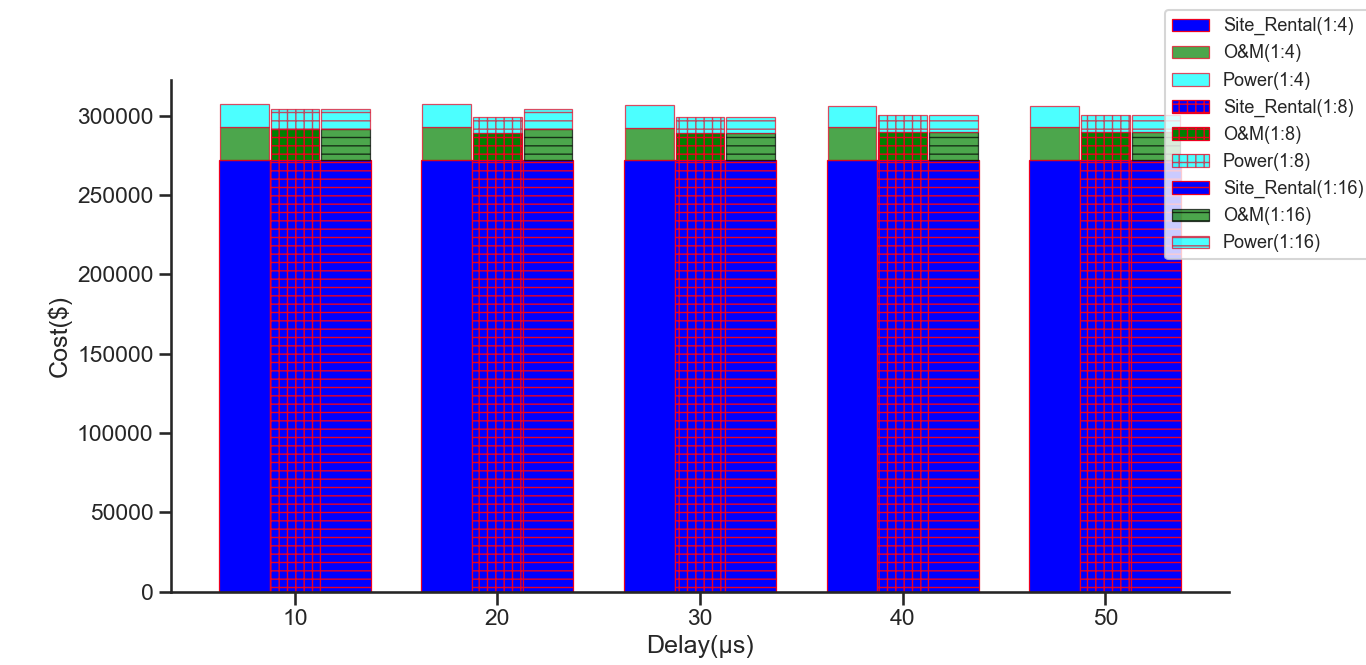}
\caption{Opex breakdown vs. delay threshold}
\label{fig:5}
\end{figure}

\begin{figure}[!htb]
\centering
\includegraphics[width=0.38\textwidth]{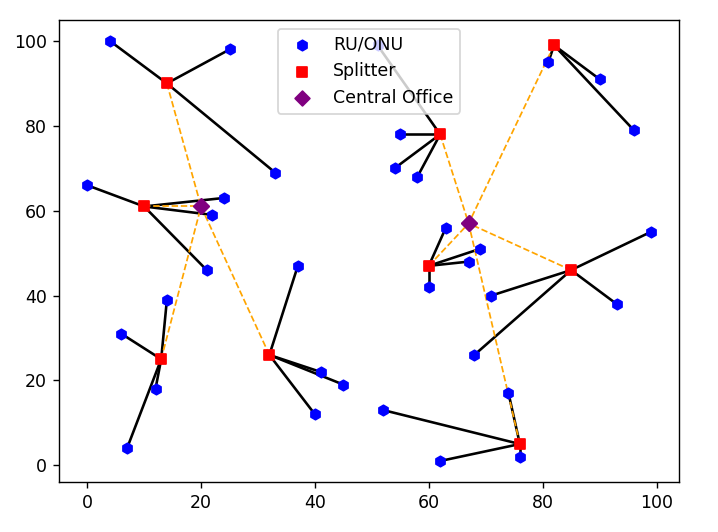}
\caption{Optimal deployment for 30 $\mu s$ with 1:4 splitting ratio}
\label{fig:6}
\end{figure}

\begin{figure}[!htb]
\centering
\includegraphics[ width=0.38\textwidth]{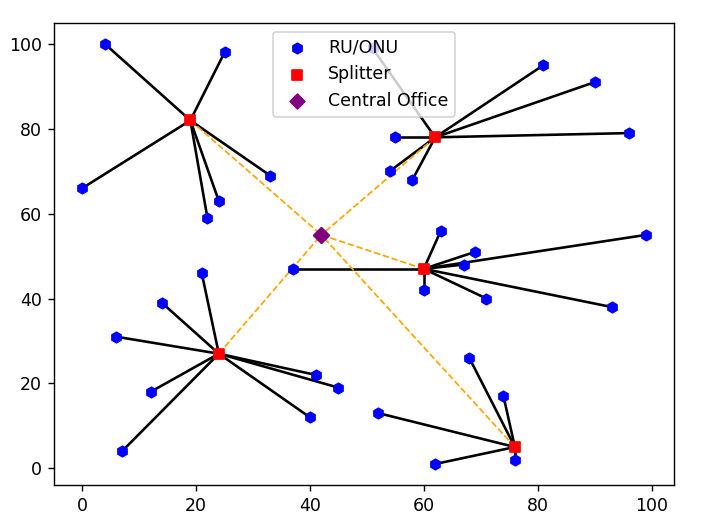}
\caption{Optimal deployment for 30 $\mu s$ with 1:8 splitting ratio}
\label{fig:7}
\end{figure}

\begin{figure}[!htb]
\centering
\includegraphics[width=0.38\textwidth]{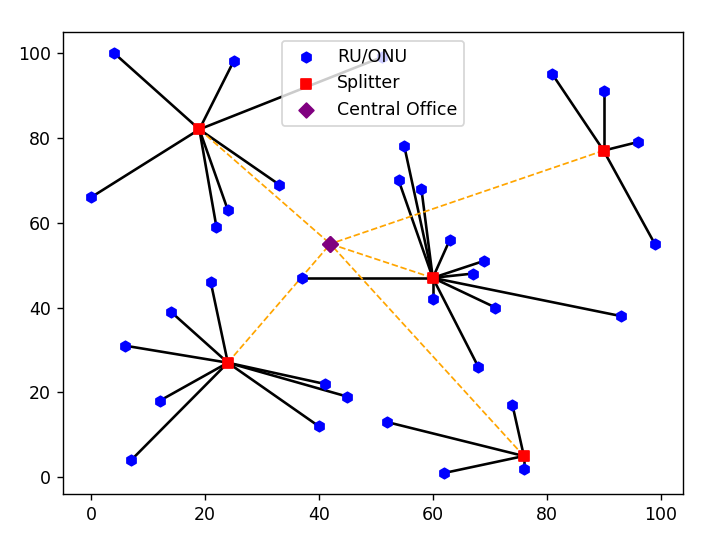}
\caption{Optimal deployment for 30 $\mu s$ with 1:16 splitting ratio}
\label{fig:8}
\end{figure}

\hspace{-0.3cm}In this section, the numerical results are obtained by scattering 34 RU/ONUs in a 10x10 $km^{2}$ greenfield area. We use the commercially available ILOG CPLEX  software to find the optimal deployment solution using a computer with 8GB RAM and  Intel i5 processor. The obtained results are compared for different delay thresholds (10, 20, 30, 40, 50) $\mu s$ and different splitting ratios (1:4, 1:8, 1:16).
Basically, 40 Gb/s \mbox{TWDM-PON} with the splitting ratios mentioned above can support functional split options ranging from 1 to 6, 7.2, and 7.3\cite{b15}. For more details about functional split options for 5G, the reader is referred to \cite{b15}. To represent the objective function, we take into account a variety of costs listed in Table~\ref{tab:3}, where the cost column of each piece of equipment includes material and installation costs. We assume that the capacity supported by each RU/ONU can be 2.5 Gb/s.  Furthermore, the capacity provided by TWDM-PON is 40 Gb/s both for uplink and downlink, respectively (symmetrical network). The maximum splitting ratio is 1:16, as 40 Gb/s TWDM-PON can not serve more than 16 RU/ONUs with capacity of 2.5 Gb/s. We assume that the maximum number of DOs hosted in one central office is 10.
We wanted to determine how the optimal cost varies with the delay threshold while meeting 5G requirements. Figure~\ref{fig:3} illustrates the optimal costs of the network for different delay thresholds. It is clear that the deployment cost decreases with the increase of delay requirements. Furthermore, the TCO  decreases as the split ratio increases, where TCO can be divided into Capex (colored blue) and Opex (colored red), and the proposed ILP can provide the TCO costs in detail.  Figure~\ref{fig:4}  represents a division of Capex for different delay thresholds and splitting ratios. There is an inverse relationship between the delay threshold and the Capex value. It is clear that  when the split ratio increases, the number
of PONs required as a fronthaul for the total number of RU/ONUs
will decrease. As a result, the numbers of splitters,
AWGs, and DOs are decreased, which reduces the total deployment
cost. On the other hand, the fiber cost of the fronthaul decreases when the split ratio decreases. A high number of equipment units occurring when the splitting ratio decreases means that the Opex value increases as shown in  Fig.~\ref{fig:5}.  We can conclude that the site rental costs are higher than O$\&$M costs and power consumption costs. 
Figs.~\ref{fig:6}-\ref{fig:8} illustrate the examples of fronthaul deployment options provided by our optimization methodology for the tested region where the delay threshold is 30 $\mu s$, and the splitting ratios are 1:4, 1:8, and 1:16, respectively. The blue dots refer to RU/ONU locations, the red squares indicate the optimal splitter locations, and the purple lozenges indicate the optimal central office locations. The feeder fiber connections are represented by orange dashed lines, while black lines represent the distribution fiber connections.
\begin{table}
\caption{Capex and Opex of the Case Study\cite{b12,b20}}
\label{tab:3}       
\begin{tabular}{p{4cm}|p{3cm}}
\hline\noalign{\smallskip}
Parameter & Cost [$\$$]  \\
\noalign{\smallskip}\hline\noalign{\smallskip}
CO housing & 75000 \\
DO & 6500 \\
AWG&250\\
Splitter 1:4&30\\
Splitter 1:8&50\\
Splitter 1:16&100\\
Fiber cable/m &20  \\
RU/ONU&3500\\
Yearly cell site rent&8000\\
Electric consumption [kWh]&0.15 \\
Operation and maintenance &10$\%$ of equipment\\
\hline
\hline
Component& Energy consumption [Wh]\\
\hline
DO  &255\\
Cooling system &500\\
RU/ONU&104\\
\noalign{\smallskip}\hline
\end{tabular}
\end{table}

\section{Conclusion}
Planning 5G networks to be ready for the upcoming time-sensitive services in a cost-effective way is vital for the operators.
 In this paper, we proposed an optimization framework for cost-effective planning of 5G fronthaul employing TWDM-PON for different delay thresholds and various splitting ratios. We demonstrated the suitability of our framework by using it to plan a 10x10 $km^{2}$ area with 34 RU/ONUs and different delay thresholds (10, 20, 30, 40, 50) $\mu s$ and three splitting ratios (1:4, 1:8, 1:16). We have shown that the delay threshold and splitting ratio have an important impact on the cost of the network, where the higher the delay, the lower the costs, and the lower the splitting ratio, the higher the costs. The proposed ILP model can provide the operators with detailed deployment costs for the data sets they provided, based on the required delay value and splitting ratio, as well as  minimize the deployment cost by determining the best fiber routes,
 the optimal locations for splitters and central offices. Because the ILP formula does not scale well as the number of RU/ONUs grows, the ongoing work includes the creation of a heuristic scheme to overcome this limitation.
\section*{Acknowledgement}
This work was supported by the CHIST-ERA grant \mbox{SAMBAS} (CHIST-ERA-20-SICT-003) funded by FWO, ANR, NKFIH and UKRI.

\end{document}